\pdfoutput=1
\documentclass{appolb}

\usepackage{amsmath,amssymb,bbm,color}
\usepackage{epsfig}
\usepackage{graphics}
\usepackage{euscript}
\usepackage{pslatex}
\usepackage{hyperref}
\usepackage{xcolor}
\usepackage{fancybox}
\usepackage{enumerate}
\usepackage{subfig}

\begin{document}

\title{
\vspace{-5.cm}
\begin{flushright}
        {\small {\bf IFJPAN-IV-2013-17}}   \\
        {\small {\bf SMU-HEP-13-23}}  \\
\end{flushright}
\vspace{0.8cm}
{\bf\Large NLO corrections in the initial-state\\ parton shower Monte Carlo}
\thanks{
This work is partly supported by 
 the Polish National Science Centre grant UMO-2012/04/M/ST2/00240,
  the Research Executive Agency (REA) of the European Union 
  Grant PITN-GA-2010-264564 (LHCPhenoNet),
the U.S.\ Department of Energy
under grant DE-FG02-13ER41996 and the Lightner-Sams Foundation.
}}

\author{S.\ Jadach$^a$, A.~Kusina$^b$, W.\ P\l{}aczek$^c$, M.\ Skrzypek$^a$
\address{$^a$ Institute of Nuclear Physics, Polish Academy of Sciences,\\
              ul.\ Radzikowskiego 152, 31-342 Krak\'ow, Poland}
\address{$^b$ Southern Methodist University, Dallas, TX 75275, USA}
\address{$^c$ Marian Smoluchowski Institute of Physics, Jagiellonian University,\\
              ul. Reymonta 4, 30-059 Krak\'ow, Poland.}
}


\maketitle

\begin{abstract}
The decade-old technique of combining NLO-corrected hard process
with LO-level parton shower Monte Carlo
is now mature and used in practice of the QCD calculations
in the LHC data analysis.
The next step,
its extension to an NNLO-corrected hard process combined
with the NLO-level parton shower Monte Carlo, will require
development of the latter component.
It does not exist yet in a complete form.
In this note we describe recent progress in developing
the NLO parton shower for the initial-state hadron beams.
The technique of adding NLO corrections in the fully exclusive form
(defined in recent years)
is now simplified and tested numerically,
albeit for a limited set of NLO diagrams in the evolution kernels.
\end{abstract}



\section{Introduction}

Perturbative Quantum Chromodynamics
(pQCD)~\cite{GWP,Gross:1974cs,Georgi:1951sr}
is the basic and indispensable tool
for analyzing experimental data in the LHC experiments.
The technique of combining an NLO-corrected hard process
with a LO-level parton shower Monte Carlo (replacing collinear PDFs),
like MC@NLO  \cite{Frixione:2002ik}
and POWHEG \cite{Nason:2004rx,Frixione:2007vw},
is now used in practice of the QCD calculations
in the LHC data analysis.
Its logical extension, providing
higher-precision QCD predictions,
would be an NNLO-corrected hard process combined
with the NLO-level parton shower Monte Carlo (MC).
However,  the NLO-level parton shower Monte Carlo does not exist yet.
In addition, the methods of NLO-correcting the hard process used
in the above methodologies are quite complicated and it
would be desirable to simplify them before going to the NNLO level.
The authors of this note are developing solutions to both above problems.
On the one hand, 
in refs.~\cite{Jadach:2011cr,Jadach:2012vs},
see also refs.~\cite{Skrzypek:2011zw,Jadach:2012sh},
they are developing a simpler method of introducing the NLO corrections
to the hard process.
On the other hand, completely new techniques of NLO-correcting
parton shower MC are developed,
see refs.~\cite{Jadach:2009gm,Jadach:2010aa}.

In the present note we show that the technique used to simplify
and speedup inclusion of the NLO corrections in the hard
process, \cite{Jadach:2011cr,Jadach:2012vs},
can also be applied for the same purpose in the methods
of refs.~\cite{Jadach:2009gm,Jadach:2010aa} to introduce
the NLO corrections in the parton shower MC.
Some similarities (and differences)
to the POWHEG \cite{Nason:2004rx} method are discussed
in refs.~\cite{Jadach:2012vs,Jadach:2012sh}, in the case of the hard process.


\section{Overview of method of NLO-correcting parton shower MC}
For the detailed description of
the methodology of NLO-correcting the parton shower MC
we refer the reader to refs.~\cite{Jadach:2010aa,Skrzypek:2011zw}.
Ref.~\cite{Jadach:2009gm} presents an older variant of the method --
on the other hand, it provides many details of the differential cross
sections of the NLO corrections to the ladder.
The above studies and this work, are limited to the non-singlet
component of the QCD evolution of the quark distributions in the hadron beam,
using non-running $\alpha_S$.
The DGLAP evolution equation is solved {\em exactly} using a simple Markovian MC
with the relevant inclusive LO or LO+NLO evolution kernels.
\footnote{%
 We shall refer to this calculation
 as an {\em ``inclusive benchmark MC''}.
 See ref.\cite{Jadach:2012vs} for details.}.
The newly developed methods use fully exclusive (unintegrated)
evolution kernels and their results, 
at the inclusive level (evolved quark $x$-distributions),
are compared with the exact inclusive MC calculation.

The algebraic structure of the NLO-corrected exclusive distributions
of the simplified parton shower MC reads as follows%
\footnote{This is eq.~(1) in ref.~\cite{Jadach:2010aa}.}
\begin{equation}
\label{eq:one}
\begin{split}
&
\rho_n(k_l)=
e^{-S_{_{ISR}}}
\Bigg\{
\raisebox{-30pt}{\includegraphics[height=30mm]{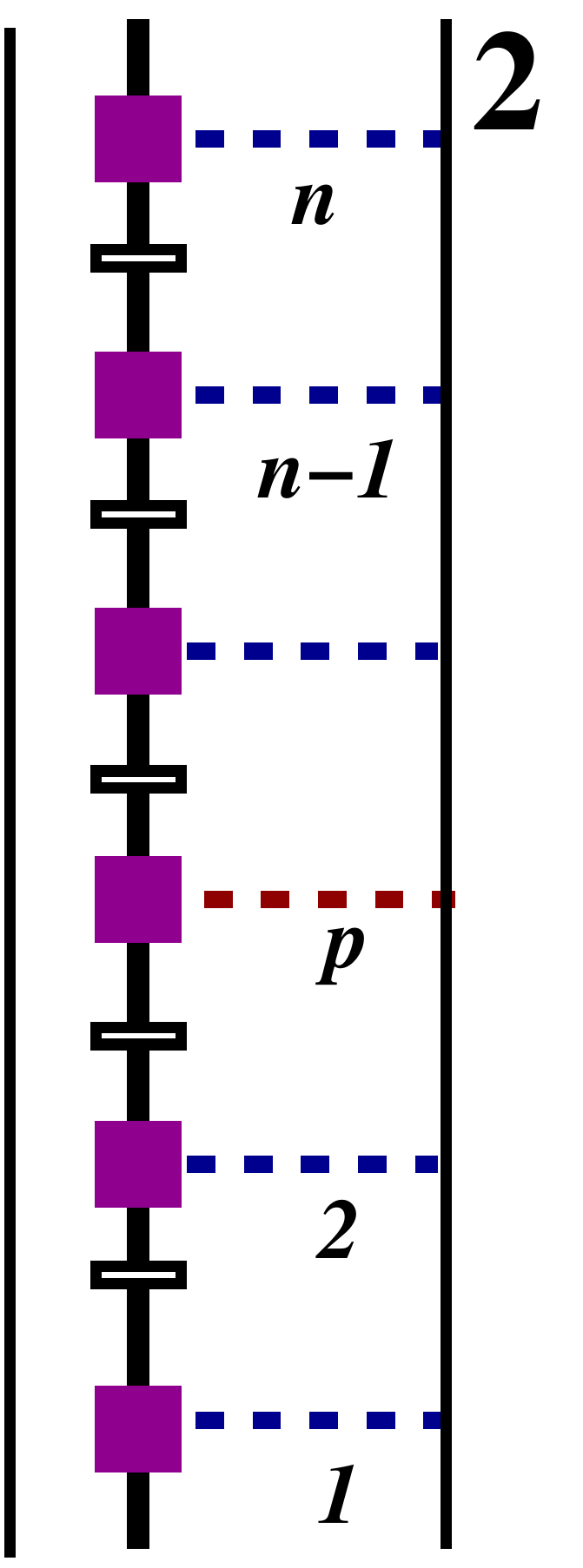}}
+
\sum\limits_{p_1=1}^{n}
\sum\limits_{j_1=1}^{p_1-1}
\raisebox{-30pt}{\includegraphics[height=30mm]{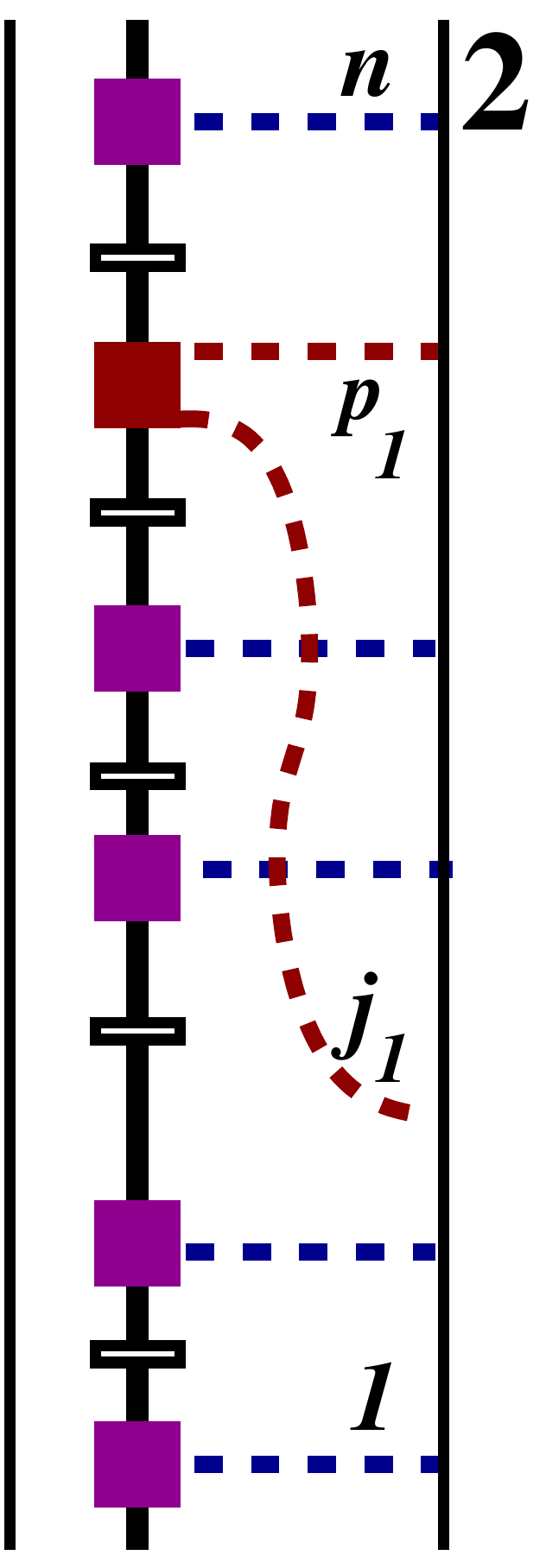}}
+
\sum\limits_{p_1=1}^{n}
\sum\limits_{p_2=1}^{p_1-1}
\sum\limits_{j_1=1 \atop j_1\neq p_2}^{p_1-1}
\sum\limits_{j_2=1 \atop j_2\neq p_1,j_2}^{p_2-1}
\raisebox{-30pt}{\includegraphics[height=30mm]{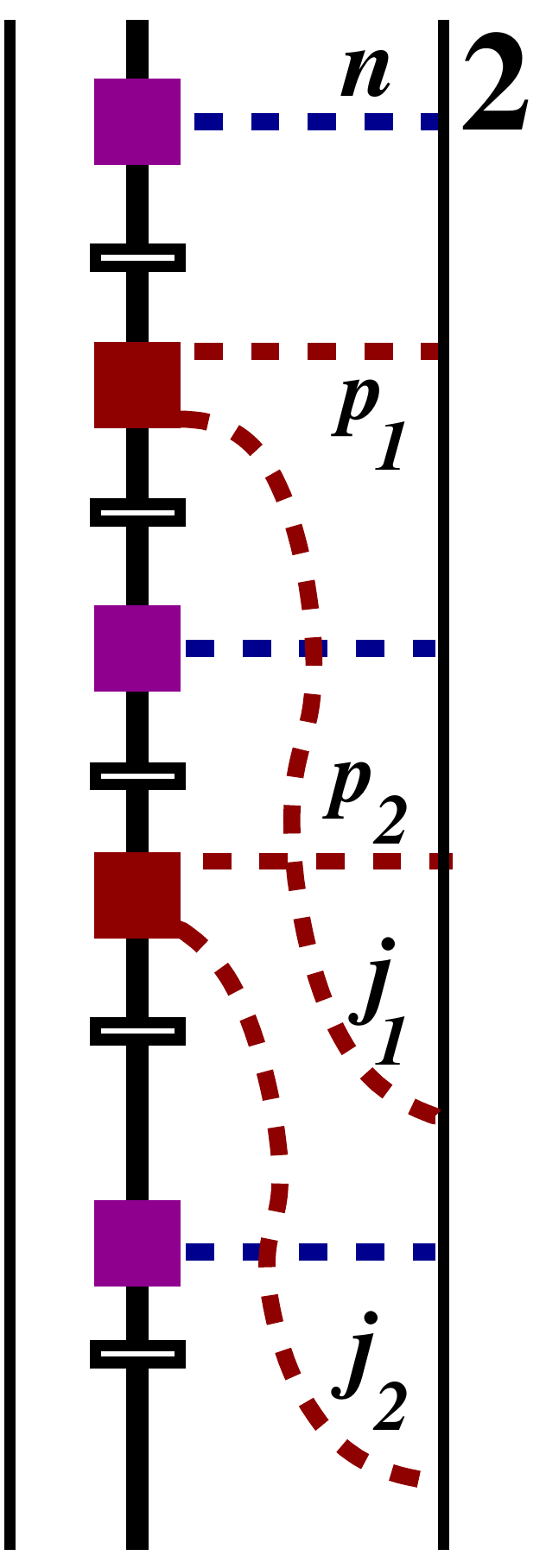}}
\Bigg\}=
\\&~~~
=e^{-S_{_{ISR}}}
\bigg[
\beta_0^{(1)}(z_p)
+ \sum\limits_{p=1}^{n} \sum_{j=1}^{p-1}W(\tilde{k}_p, \tilde{k}_j)+
\\&~~~~~~~
+\sum\limits_{p_1=1}^{n}
\sum\limits_{p_2=1}^{p_1-1}
\sum\limits_{j_1=1 \atop j_1\neq p_2}^{p_1-1}
\sum\limits_{j_2=1 \atop j_2\neq p_1,j_2}^{p_2-1}
\!\!\!
W(\tilde{k}_{p_1}, \tilde{k}_{j_1})
W(\tilde{k}_{p_2}, \tilde{k}_{j_2})
+\dots
\bigg]
\prod_{i=1}^n\; \theta_{a_i>a_{i-1}}
    \rho^{(1)}_{1}(k_i)
    \beta_0^{(1)}(z_i).
\end{split}
\end{equation}
Notation and definitions can be found in ref.~\cite{Jadach:2010aa}.
For the purpose of the following discussion let us only recall the
definition of the weight $W$. It introduces the 2-real NLO correction
involving $C_F^2$ subtracted part of the exact matrix element for the emission
of two gluons from the quark line (including interference):
\begin{equation}
W(k_2,k_1)=
\frac{%
 \left| \raisebox{-8pt}{\includegraphics[height=8mm]{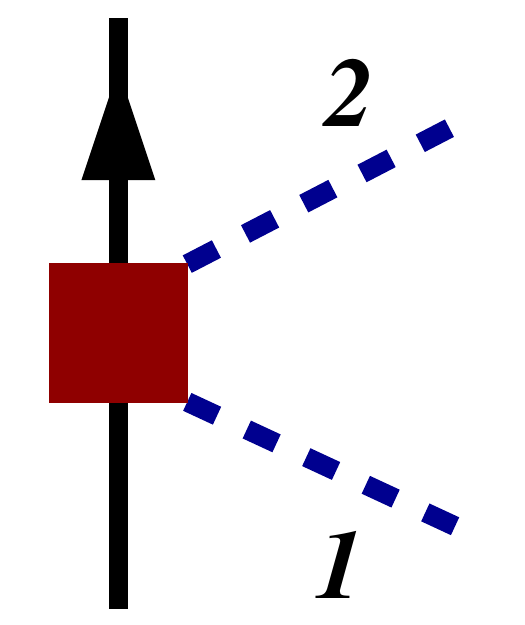}}
 \right|^2
}{%
 \left| \raisebox{-8pt}{\includegraphics[height=8mm]{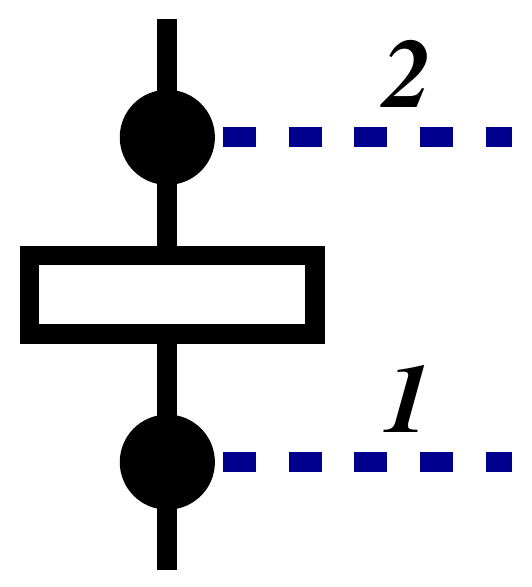}}
 \right|^2
}=
\frac{%
 \left| \raisebox{-8pt}{\includegraphics[height=8mm]{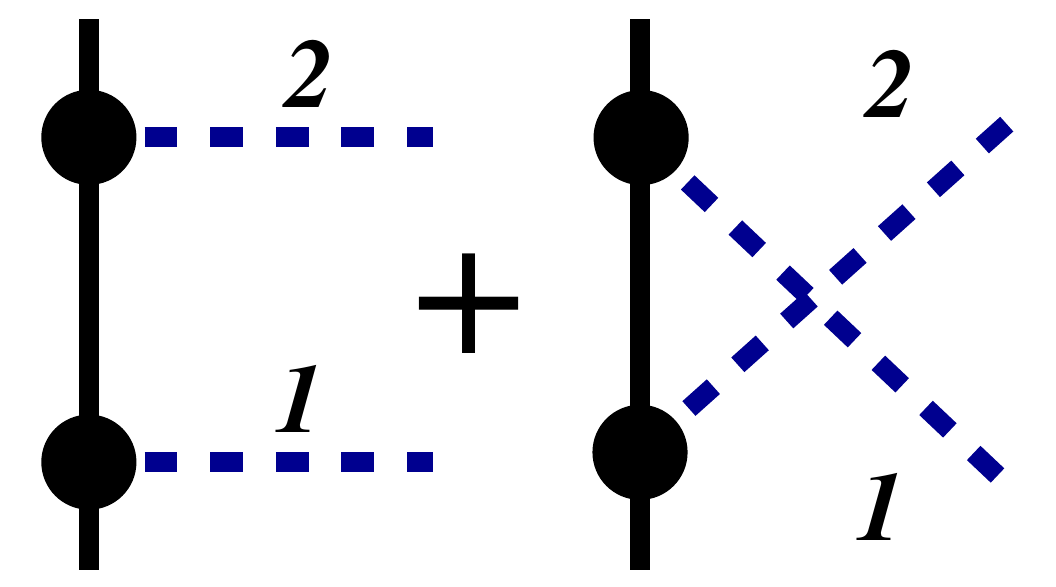}}
\right|^2
}{%
 \left| \raisebox{-8pt}{\includegraphics[height=8mm]{xBr2ReCt.pdf}}
 \right|^2
}\; -1
\end{equation}
The other, triple, vertex aggregates the LO kernel with
all unresolved (virtual+soft) corrections (excluding the Sudakov part)%
\footnote{%
   See also ref.~\cite{gituliarustron2013}.}:
\begin{equation}
\left|
 \raisebox{-10pt}{\includegraphics[height=10mm]{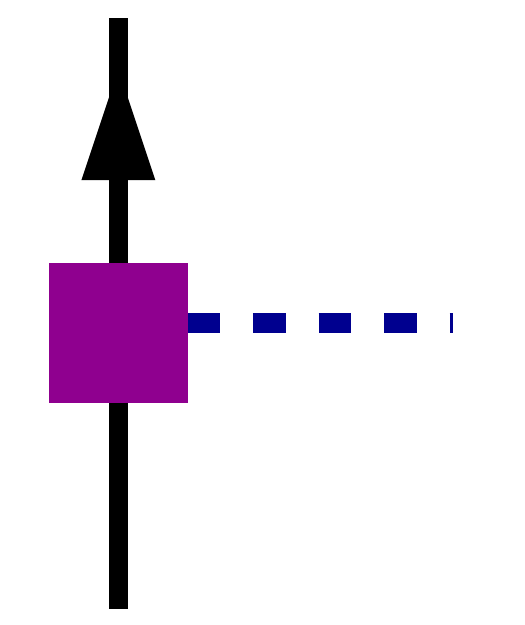}}
\right|^2\!\!\!
=\big(1+2\Re(\Delta_{_{ISR}}^{(1)})\big)\!
\left|
  \raisebox{-10pt}{\includegraphics[height=10mm]{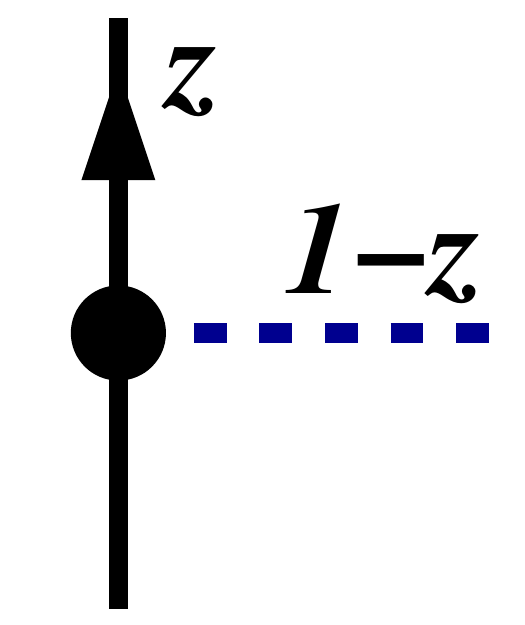}}
\right|^2.
\end{equation}
It is very important that
both the above building blocks of the NLO corrections
are free of any infrared or collinear singularities.

In eq.~(\ref{eq:one}) the summations over indices
$p_1$ and $p_2$
are over the positions of the so-called ``NLO insertions'',
which upgrade one and two kernels to the NLO level.
The triple and higher order summations, which are upgrading three
and more kernels could be included, but we have checked that
they are numerically unimportant.
On the other hand, summations over ``spectator gluons''
$j_1$ and $j_2$ are important and they are regarded as a landmark
of our method.
(They are similar to the sums over $\tilde\beta$ non-infrared terms
in the QED exponentiation scheme of ref.~\cite{Jadach:2000ir}.)
These sums may slow down the generation of the MC events
and are rendering the evaluation of the MC weight quite complicated.

\begin{figure*}[!ht]
  \centering
  {\includegraphics[width=0.8\textwidth]{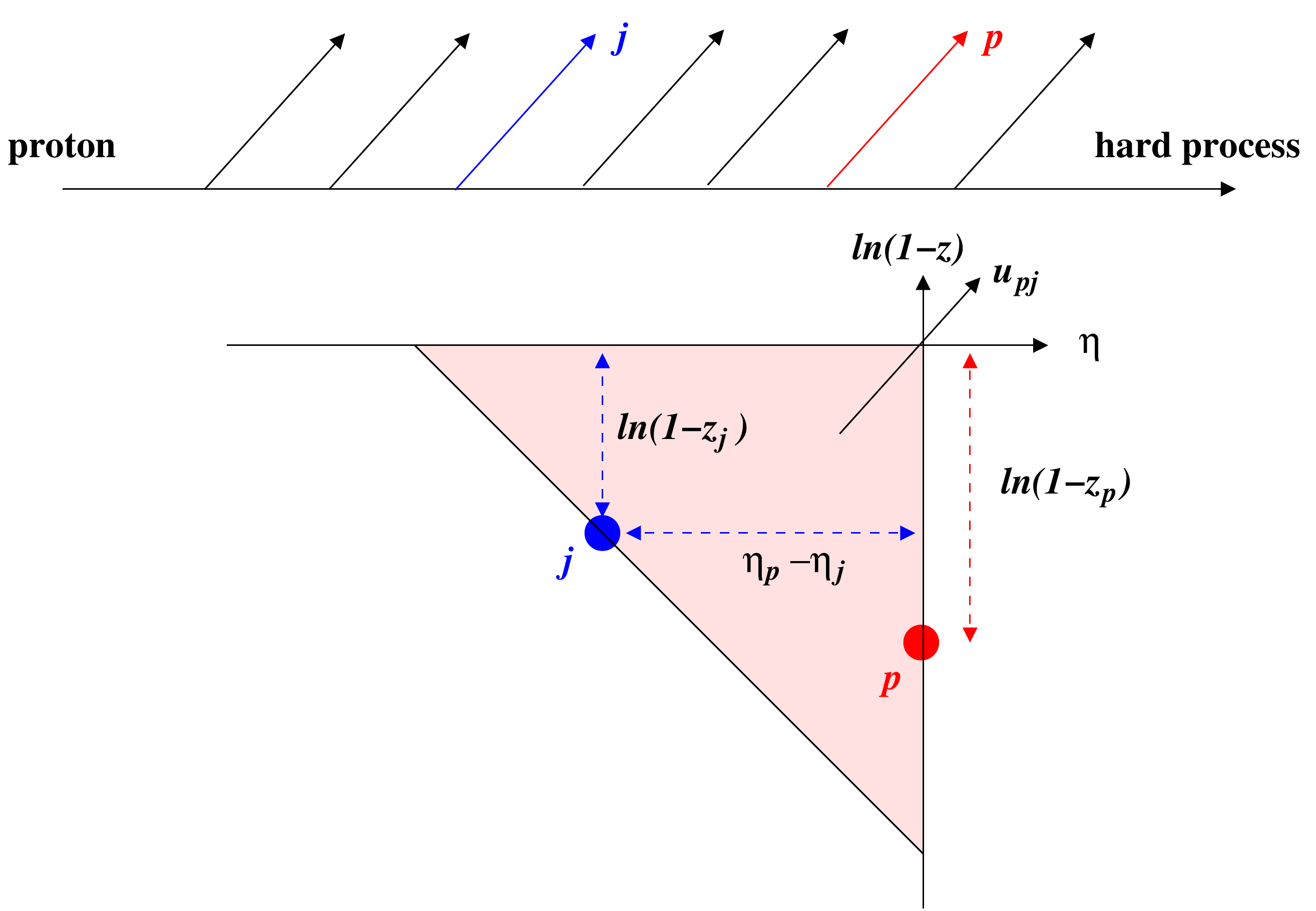}}
  \caption{
    The kinematics of the two-gluon phase space of the NLO correction.
    The variable $\eta$ is rapidity of the emitted gluon
    and $z$ is the conventional lightcone variable of the emitter quark.
    }
  \label{fig:kinema}
\end{figure*}

In refs.~\cite{Jadach:2012vs,Jadach:2012sh} it was shown how to
reduce the sums $j_i$ over spectator gluons just to one or two terms,
limiting these sums to contributions from one or two
gluons with maximum transverse momentum%
\footnote{Similarly as in POWHEG, but without complicated ``vetoed''
 and ``truncated'' MC showers.},
without loosing the completeness of the NLO approximation.
Hence, it is obvious to ask whether a similar ``trick'' is possible here,
in eq.~(\ref{eq:one}).
The key point is to invent within the ladder kinematics
some new variable which could be used to define easily a spectator gluon as
the hardest one -- the only one which ``saturates'' the sum $j_i$
over spectators%
\footnote{In the sense of protecting completeness of the NLO.}.
We cannot use directly $k^T_j$ of the spectator,
because the phase space of the NLO
correction is really the two-gluon phase space --
a new variable $u_{pj}$ has to involve
momenta of both gluons, the ``head'' $p$ and the ``spectator'' $j$.
Moreover, similarly as $k^T$ in the hard process, it has to
provide the ``Sudakov suppression'' in the limit $u_{pj}\to 0$.

In Fig.~\ref{fig:kinema} we illustrate the problem and the solution
in a graphical way. The solution is the following:
\begin{equation}
u_{pj}  = \eta_p-\eta_j +\lambda\ln(1-z_j).
\end{equation}
The paremeter $\lambda\simeq 1$ will provide an extra optimization
in the following numerical exercises.
The direction of $u_{pj}$ is marked in Fig.~\ref{fig:kinema} --
it points towards the tip of the shaded triangle which marks the endpoint
of the allowed phase space of the spectator gluon $j$.
In the essence variable $\exp(u_{pj})$ represents the rescaled $k^T$
of the spectator gluon $j$.
The above kinematics describes a parton shower MC with the angular ordering,
however, the kinematics of the parton shower with 
the $k^T$-ordering is quite similar.

\begin{figure*}[h]
  \centering
  {\includegraphics[width=1.0\textwidth]{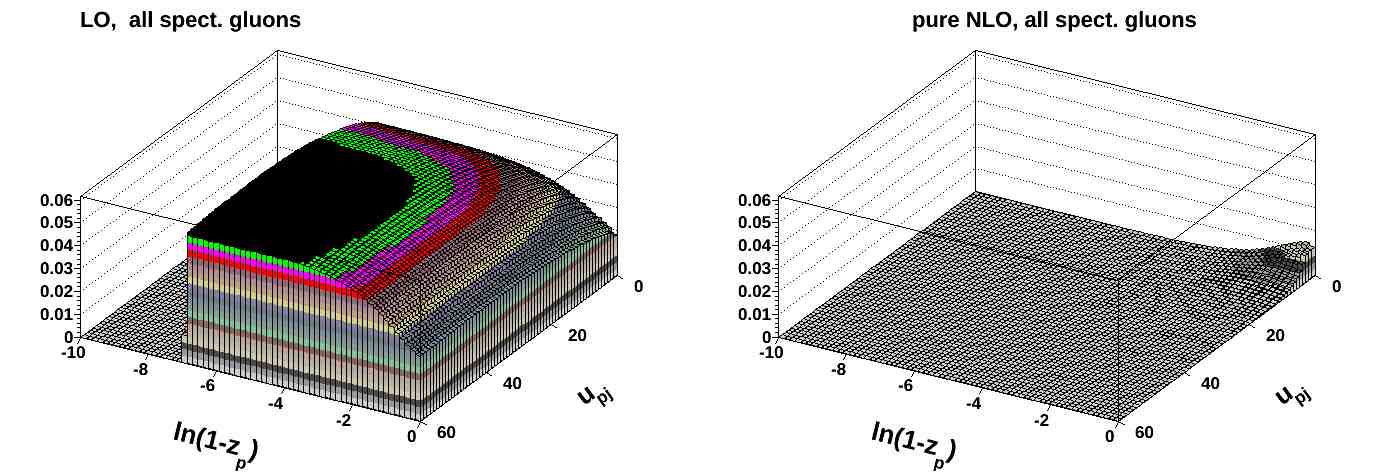}}
  \caption{
    The inclusive distribution of gluons according to LO distribution (left)
    and due to the two-real NLO contribution (right).
    }
  \label{fig:two_gluon}
\end{figure*}

In the above double-gluon phase space with fixed rapidity of the head
gluon $p$, the Sudakov phase space is 3-dimensional,
($\eta_p-\eta_j, \ln(1-z_j), \ln(1-z_p)$),
and the volume of the 
underlying 3-dimensional LO gluon phase space is equal to triple Sudakov log.
In the 2-dimensional
visualization of this phase space gluon density in Fig.~\ref{fig:two_gluon}
we use a set of variables  $\ln(1-z_p)$ and $u^2_{pj}$ in order to have a flat
plateau representing manifestly the leading LO Sudakov singularity.
The LHS of Fig.~\ref{fig:two_gluon} shows this Sudakov LO plateau.

On the other hand, the NLO contribution plotted in the RHS of Fig.~\ref{fig:two_gluon}
clearly concentrates in the corner $z_p\simeq 0, u_{pj} \simeq 0$,
and is manifestly free of any singularities
(it is integrable to a finite value).
This is quite similar as in the single-gluon phase space of
the hard process shown in Fig.~5 in ref.~\cite{Jadach:2012vs}.

\begin{figure*}[!ht]
  \centering
  {\includegraphics[width=1.0\textwidth]{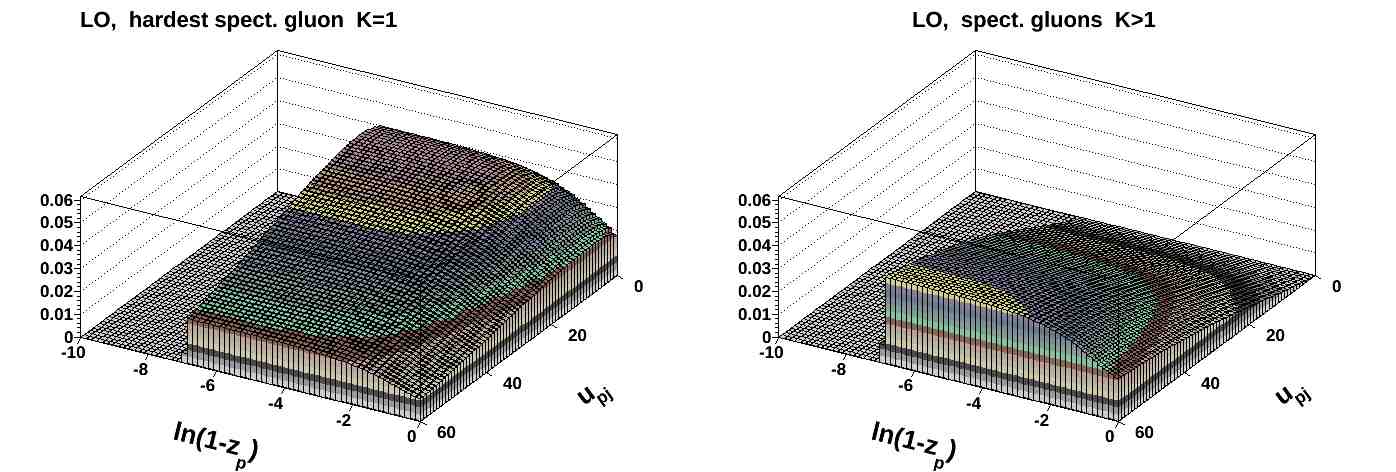}}
  \caption{
    The inclusive LO distribution of gluons of 
    the left plot in Fig.~\ref{fig:two_gluon}
    split into the hardest (in $u_{pj}$) gluon (left) and the rest (right).
    }
  \label{fig:LOsplit}
\end{figure*}
In the next step, let us order spectator gluons (indexed by $j$)
and split the LO distribution 
(similarly as in Fig.~6 of ref.~\cite{Jadach:2012vs})
into the hardest in the variable $u_{pj}$ and the rest%
\footnote{We cannot order in $\ln(1-z_p)$ 
  because the head gluon $p$ is just one.}.
The resulting two components are shown in Fig.~\ref{fig:LOsplit}.
The hardest gluon distribution differs from the one in 
Fig.~6 of ref.~\cite{Jadach:2012vs}, nevertheless it has the same property
needed for NLO completeness -- it reproduces well the inclusive LO
distribution (LHS plot in Fig.~\ref{fig:two_gluon}) in the region where
the NLO contribution (RHS plot in Fig.~\ref{fig:two_gluon}) is nonzero.

In view of the above,
we expect that preserving
only one term in the sums over $j$ in eq.~(\ref{eq:one}),
from the gluon with the maximum $u_{pj}$,
will effectively lead to NLO result within a good numerical approximation
(formally up to NNLO terms).
We shall check this conjecture in the following.

\begin{figure*}[!ht]
  \centering
  {\includegraphics[width=1.0\textwidth,height=110mm]{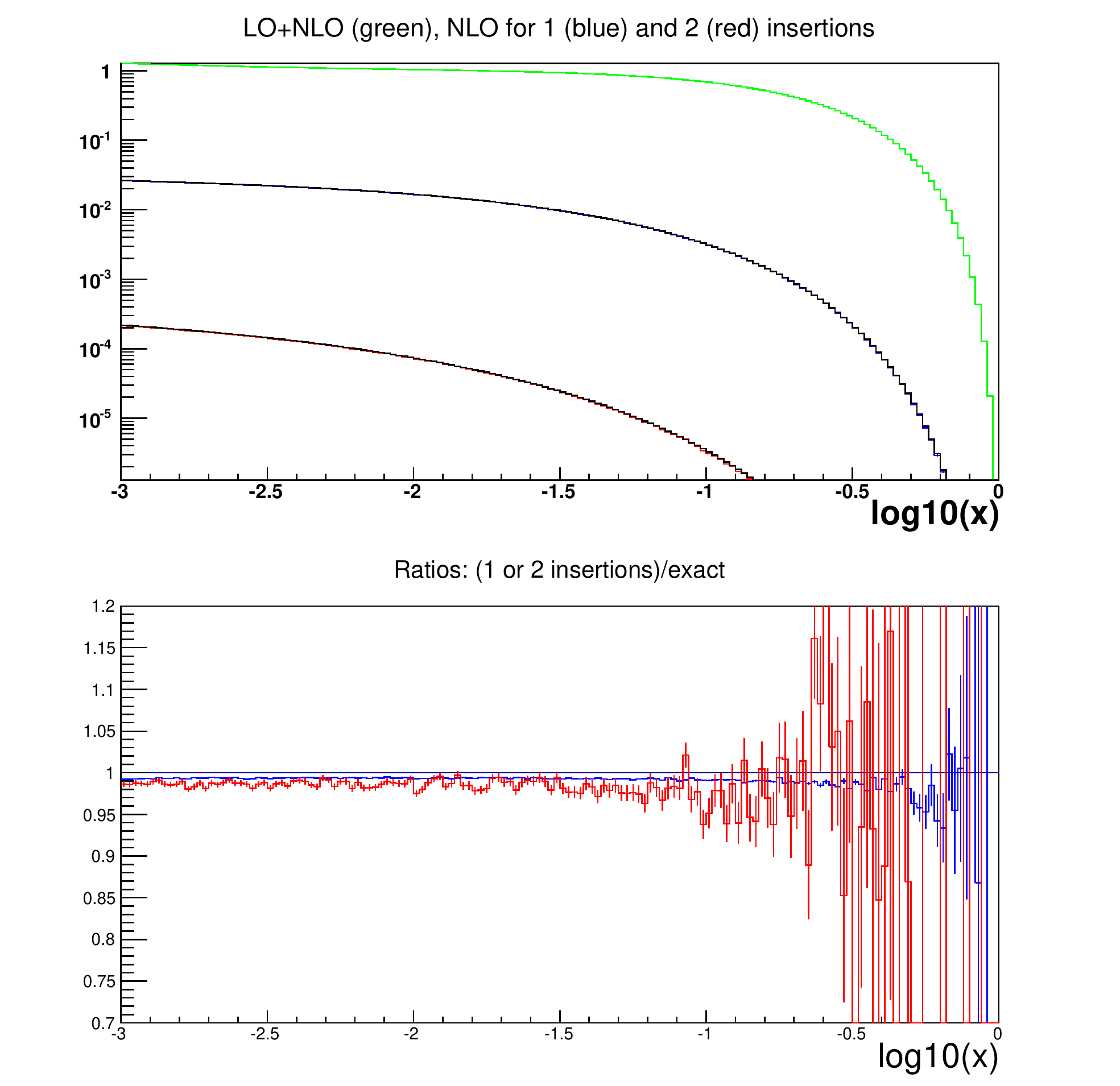}}
  \caption{
The distribution of the quark evolved from $Q=100$~GeV to $Q=10$~TeV.
In the upper plot the upper line represents LO+NLO quark distribution 
and two lines
below are the components due to 1 and 2 NLO insertions in eq.~(\ref{eq:one}).
The corresponding inclusive benchmark results are also plotted, but they are
indistinguishable, hence the corresponding ratios (exclusive/inclusive) are
provided in the lower plot.
    }
  \label{fig:oldresult}
\end{figure*}

\section{Numerical results}
In the following we shall check numerically
that taking only one or two hardest (in $u$-variable) spectator gluons
in the NLO MC weight of eq.~(\ref{eq:one}) does not significantly disturb
the NLO result of the QCD evolution.
This is the principal result of this work.


As a warm-up exercise we reproduce the result of ref.~\cite{Jadach:2012sh},
in which we use eq.~(\ref{eq:one}) with summation over 
all spectator gluons $j_1$ and $j_2$.
In Fig.~\ref{fig:oldresult} the total (LO+NLO) 
quark distribution evolved with single and
double NLO insertion is compared with the benchmark inclusive calculation.
The two are indistinguishable, and to see the difference one should look
at the lower plot in Fig.~\ref{fig:oldresult},
where the ratios of the exclusive and inclusive results are plotted
for the single and double NLO insertions separately.
They agree perfectly within the statistical errors.

\begin{figure*}[!ht]
  \centering
  {\includegraphics[width=1.0\textwidth,height=110mm]{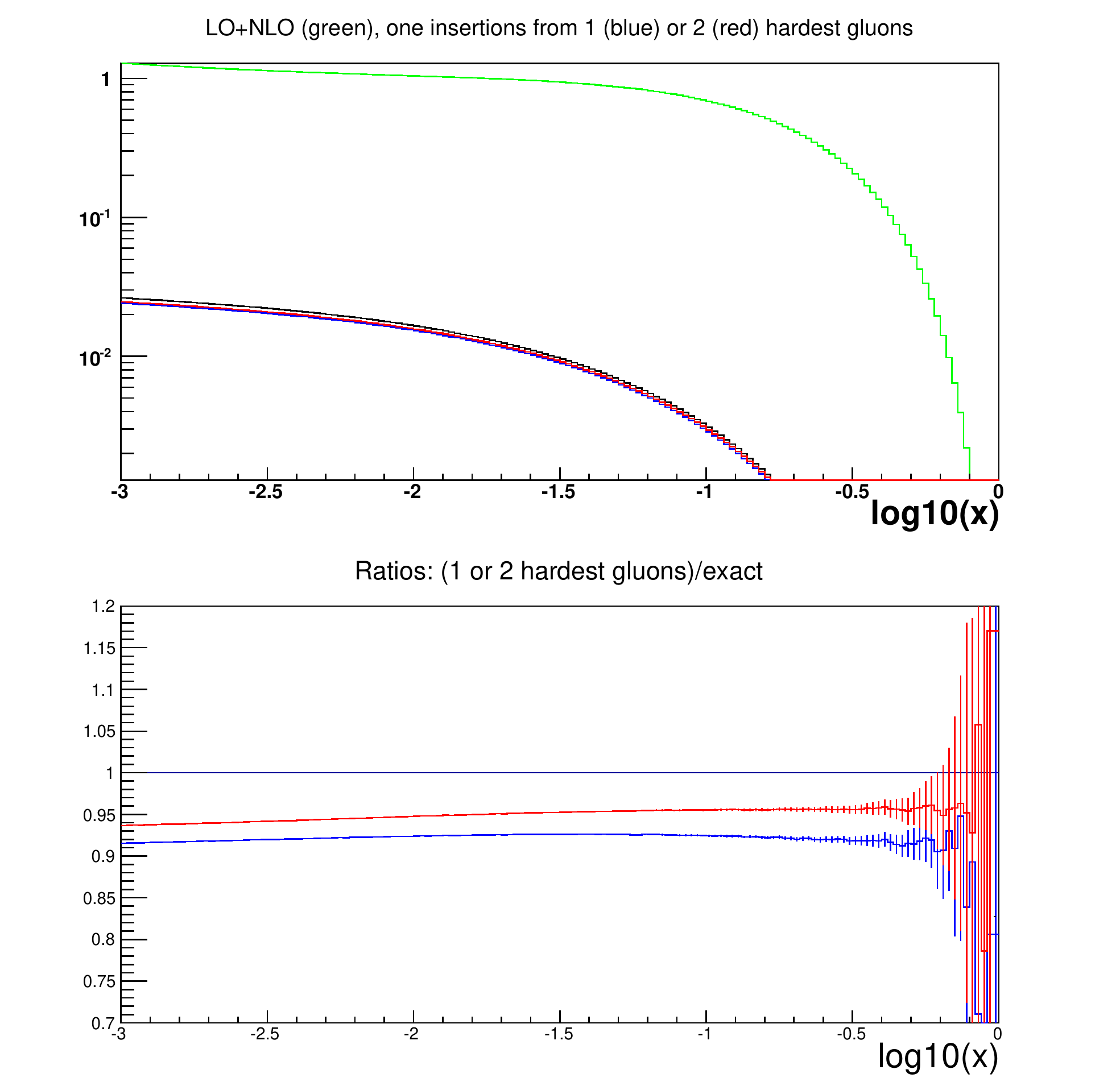}}
  \caption{
    The upper line represents LO+NLO quark distribution and three lines below
    represent a single NLO insertion component
    with the complete sum over spectator gluons as in eq.~(\ref{eq:one})
    and two other versions with the sum truncated to one and two hardest
    spectator gluons.
    The two corresponding ratios (truncated/complete) are shown in
    the lower plot.
    }
  \label{fig:newresult}
\end{figure*}

Next, in the calculation presented in
Fig.~\ref{fig:newresult}, we replace the sums over spectators
with the one or two terms from the hardest gluons in the variable $u_{pj}$,
for the single gluon insertion component.
As we see, this truncated result reproduces very well the previous
single NLO insertion component in the evolved quark distribution.
The actual difference is better seen in the lower plot
of Fig.~\ref{fig:newresult} representing the ratios of the truncated
and complete sums over spectator gluons.
Of course, the case with two hardest spectator gluons looks better,
but the single hardest gluon would be sufficient.
It should be added that in the above result we have adjusted
$\lambda=2$ in the definition of $u_{pj}$.
For $\lambda=1$ the ratio for single spectator gluon would be $\sim0.7$
at the low $x$ limit
(remaining formally all the time correct modulo NNLO corrections).

\section{Summary and outlook}
A new methodology of adding 
the QCD NLO corrections to the NLO 
initial state Monte Carlo parton shower
is refined and tested numerically,
albeit for a limited set of the NLO diagrams
and in the simplified MC model.
This result presents another important step towards
realistic implementation of the NLO parton shower MC,
to be combined with the NNLO-corrected hard process.


\section*{Acknowledgment}
Two of the authors (S.J. and M.S.) are grateful for
the warm hospitality of the TH Unit of the CERN PH Division,
while completing part of this work.



\end{document}